\documentclass[a4paper, twocolomns, 12pt]{article}
\usepackage[left=19mm, right=19mm, bottom=33mm, top=33mm]{geometry}
\pdfoutput=1
\pdftrue

\usepackage{amsmath}
\usepackage[pdftex]{graphicx}

\usepackage[T2A]{fontenc}
\usepackage[utf8]{inputenc}
\usepackage[english]{babel}
\usepackage{multicol}
\usepackage{colortbl} 

\usepackage{geometry}
\usepackage{amsfonts}
\usepackage{amsmath}
\usepackage{amssymb}
\usepackage{ntheorem}
\usepackage{latexsym}
\usepackage{units}
\usepackage{textcomp}
\usepackage{wrapfig}
\usepackage{booktabs}
\usepackage{hyperref}

\usepackage{xcolor}
\hypersetup{
    colorlinks,
    linkcolor={red!50!black},
    citecolor={blue!50!black},
    urlcolor={blue!80!black}
}

\everymath{\displaystyle}
\usepackage{color}
\everymath{\displaystyle}

\renewcommand{\phi}{\varphi}

\usepackage{url}

\usepackage{sectsty} 
\sectionfont{\fontsize{11}{14}\selectfont}     
\subsectionfont{\fontsize{10}{13}\selectfont}  

\newcommand{\descrfnt}{\fontfamily{ppl}\selectfont\small} 

\begin{document}
~\\
~
\vspace{20mm}

\begin{multicols}{2}
\end{multicols}

\vspace{-30mm}
\section*{\Large ~~~~ A search for the dark photon in the OKA experiment}
\begin{center}
\begin{minipage}{1.0\linewidth}
  {\center \large \textsc{The OKA collaboration}\\}\vspace{-2mm}
\end{minipage}
\end{center}
\begin{center}
\begin{minipage}{1.0\linewidth}
 \center{
  A.~S.~Sadovsky{${}^{a}$},
  A.~P.~Filin,
  A.~V.~Artamonov,
  S.~V.~Donskov,
  A.~M.~Gorin,
  A.~V.~Inyakin,
  G.~V.~Khaustov,
  S.~A.~Kholodenko{${}^{b}$},
  V.~N.~Kolosov,
  A.~K.~Konoplyannikov,
  V.~F.~Kurshetsov,
  V.~A.~Lishin,
  M.~V.~Medynsky,
  A.~G.~Miagkov,
  V.~F.~Obraztsov,
  A.~V.~Okhotnikov,
  V.~I.~Romanovsky{${}^{c}$},
  V.~I.~Rykalin,
  V.~D.~Samoylenko, 
  M.~M.~Shapkin,
  S.~R.~Slabospitsky,
  A.~E.~Sobol,
  I.~S.~Tiurin,
  V.~A.~Uvarov,
  O.~P.~Yushchenko
 }\vspace{-3mm}
 \center{
   \footnotesize
   \textsc{(NRC "Kurchatov Institute"${}^{}_{}{}^{}$--${}^{}_{}{}^{}$IHEP, 142281 Protvino, Russia),} 
 }\vspace{-1mm}
 \center{
  \textsc
  S.~N.~Filippov,
  E.~N.~Gushchin,
  A.~A.~Khudyakov,
  V.~I.~Kravtsov,\\
  Yu.~G.~Kudenko{${}^{d,e}$},
  A.~V.~Kulik,
  A.~Yu.~Polyarush{$^{\dagger}$},
 }\vspace{-3mm}
 \center{
   \footnotesize
   \textsc{(Institute for Nuclear Research -- Russian Academy of Sciences, 117312 Moscow, Russia),} 
 }\vspace{-1mm}
 \center{
  \textsc
  V.~N.~Bychkov, 
  G.~D.~Kekelidze,
  V.~M.~Lysan,
  V.~A.~Polyakov,
  B.~Zh.~Zalikhanov
 }\vspace{-3mm}
 \center{
   \footnotesize
   \textsc{(Joint Institute of Nuclear Research, 141980 Dubna, Russia)}\\
 }\vspace{-1mm}
\end{minipage}
\end{center}

{
\footnotesize
\line(1,0){170}\\
\vspace{-1mm}${}$\hspace{0.8cm}${}^{a}$~e-mail: Alexander.Sadovskiy@ihep.ru\\
\vspace{-1mm}${}$\hspace{0.8cm}${}^{b}$~Now at European Organization for Nuclear Research (CERN), Geneva, Switzerland\\
\vspace{-1mm}${}$\hspace{0.8cm}${}^{c}$~Now at University of Cincinnati, Cincinnati, OH, United States\\
\vspace{-1mm}${}$\hspace{0.8cm}${}^{d}$~Also at National Research Nuclear University (MEPhI), Moscow, Russia\\
\vspace{-1mm}${}$\hspace{0.8cm}${}^{e}$~Also at Moscow Institute of Physics and Technology (MIPT), Moscow, Russia\\
\vspace{-1mm}${}$\hspace{0.8cm}${}^{\dagger}$~Deceased
}
\vspace{-4mm}
\begin{center}
\begin{minipage}{0.09\linewidth}
~
\end{minipage} 
\begin{center}
\begin{minipage}{0.75\linewidth}
{ 
  \rmfamily
  {\bf Abstract}
 A search for the massless dark photon in the decay $K^{+}\to \pi^{+}\pi^{0}$\hspace{0.8pt}$\overline{{} \gamma}$ 
 is performed with the OKA detector exposed to 17.7~GeV/c RF separated secondary beam of the U70 Proton Synchrotron.
 A high-statistics data sample of the $K^{+}$ decays is used 
 for a missing mass search of the stable massless invisible dark photon ($\overline{\gamma}$) in the final state.
 In the absence of a statistically valuable signal 
 the upper limit on the branching ratio of the decay $Br<2.0\times10^{-6}$ is obtained with a 90\% confidence level based on a realistic matrix element of the process.
 Currently, it is the best upper limit for this hypothetical decay channel.
}\vspace{2mm}
\\  {\bf Keywords} {Kaon decays~$\cdot$~massless dark photon~$\cdot$~physics beyond the Standard Model~$\cdot$~dark matter: hidden sector} 
\end{minipage}
\end{center}
\begin{minipage}{0.09\linewidth}
~
\end{minipage}
\end{center}
\vspace{-2mm}


\section{Introduction}\label{sectIntro}
\vspace{-7pt}
A motivation for considering dark matter (DM) originally arose from astronomical observations that required 
an additional invisible mass to explain abnormally high velocities of peripheral stars and galaxies in large stellar formations \cite{fZwicky}. 
Although, by initial assumption, DM does not interact with an ordinary matter except through the gravitational interaction, there
still remains a possibility of a super-weak interaction between DM and SM,
see recent review \cite{mCirelli_aStrumia_jZupan_review}.
One promising approach to detect DM candidates is to hunt for the dark photons, assuming that they have a small mass or are massless.
Thus, an additional (dark) Abelian gauge symmetry group $U(1)_{D}$ is introduced \cite{Dobrescu_MasslessGaugeBosons2004}.
Depending on whether the $U(1)_{D}$ symmetry is broken or not, the associated gauge boson  -- the dark photon (DP) -- would be massive or strictly massless.
It was noted  \cite{mFabbrichesi_etal_PRL119_031801}, that the main search path for dark photons has been that of massive DP \cite{Gninenko_etal_DarkSectorNA64, aCaputo_etal_DarkPhotonHandbook, jCline_StatusDarkPhotons_CERN_TH}, 
while a strictly massless dark photon received less attention.
Given that the viable parameter space of the massive dark photon continues to shrink with accumulating null outcomes of its searches, it is of great interest to turn to the alternative possibility \cite{jySu_jTandean_EPJC80_2020_9_824}.
In that case there is no mixing of massless DP (often marked as $\overline{\gamma}$) with electromagnetic photon $\gamma$.
Hence, the interaction of $\overline{\gamma}$, with SM particles could only go through loops of massive DM particles, 
which would be the only responsible candidates for the interaction with SM particles in that case.
One of the scenarios in which $\overline{\gamma}$ could be produced with potentially observable rates involves flavor-changing-neutral-current (FCNC) decays of heavy flavors into a massless dark photon \cite{eGabrielli_etal_FCNC_decays_of_SMferm_IntoDarkPhoton},
which is considered in details in \cite{mFabbrichesi_etal_PRL119_031801, jySu_jTandean_EPJC80_2020_9_824} with particular emphasis to kaon decays. 
Namely, the $K^{+} \to \pi^{+} \pi^{0} \overline{\gamma}$ ~was proposed as a sensitive probe for massless dark photons with an estimate of branching fraction $\lesssim 1.6\times 10^{-7}$ 
assuming FCNC transition $s \to d \overline{\gamma}$ ~in a simplified model of the dark sector for rather conservative choice of parameters, where bounds from astrophysics and $K^{0}-\overline{K^{0}}$ mixing were taken into account \cite{mFabbrichesi_etal_PRL119_031801}.

The proposed FCNC approach for $\overline{\gamma}$ production was further explored for kaon and hyperon decays in \cite{jySu_jTandean_EPJC80_2020_9_824}, 
where, model-independent constraints on the decays were drawn, 
assuming that new physics connected with $\overline{\gamma}$ may be hiding in error bars of currently available branchings of hyperon decays quoted in PDG \cite{PDG}.
As a result a softer limit of $Br(K^{-}\to \pi^{-}\pi^{0}\overline{\gamma})<2.4\times10^{-6}$ was established, 
which is already within the achievable sensitivity of OKA data, cf.~our recent result on ALP \cite{OKA_AXION_EPJC}. 
In this article we perform a search for the massless dark photon.
We use eq.~(14) in \cite{jySu_jTandean_EPJC80_2020_9_824} in which an explicit matrix element for $K^{-} \to \pi^{-} \pi^{0} \overline{\gamma}$ ~is given. Using that we derived:
$$
 |M_{K^{+}\to\pi^{+}\pi^{0}\overline{\gamma}}|^{2}=-{64{a^{2}_{T}}\over{f_{\pi}^{2}}}\left[ m^{2}_{\pi^{+}} (p_{\pi^{0}}p_{\overline{\gamma}})^{2} + m^{2}_{\pi^{0}} (p_{\pi^{-}}p_{\overline{\gamma}})^{2} - 2(p_{\pi^{-}}p_{\pi^{0}}) (p_{\pi^{-}}p_{\overline{\gamma}}) (p_{\pi^{0}}p_{\overline{\gamma}})\right](|\mathbb{C}|^2 + |\mathbb{C}_{5}|^2),
$$
where $f_{\pi}=92$~MeV is the pion decay constant, $a_{T}=0.658(23)$/GeV is a constant from lattice QCD, $m_{\pi}$ is the mass of corresponding pion, 
$p_{\pi}$, $p_{\overline{\gamma}}$ are four momenta of pions and $\overline{\gamma}$,
$\mathbb{C}$ and $\mathbb{C}_{5}$ are constants  which have the dimension of inverse mass and can be complex \cite{jySu_jTandean_EPJC80_2020_9_824}.

\section{The OKA detector} 
\vspace{-5.0pt}
The OKA is a fixed target experiment dedicated
to investigation of kaon decays. It is located at NRC ''Kurchatov Institute''-$^{}$IHEP in Protvino (Russia). 
A secondary hadron beam with enhanced fraction of kaons is provided by the U-70 Proton Synchrotron using the beam line equipped with two 
CERN \cite{CERN_KaonSeparator} RF cavities operated in Panofsky scheme \cite{OkaSecondaryBeam}.
The beam has about 12.5\% of $K^{+}$ with momentum of 17.7~GeV/c.

\begin{figure}[!ht]
\centering
\includegraphics[width=0.98\textwidth]{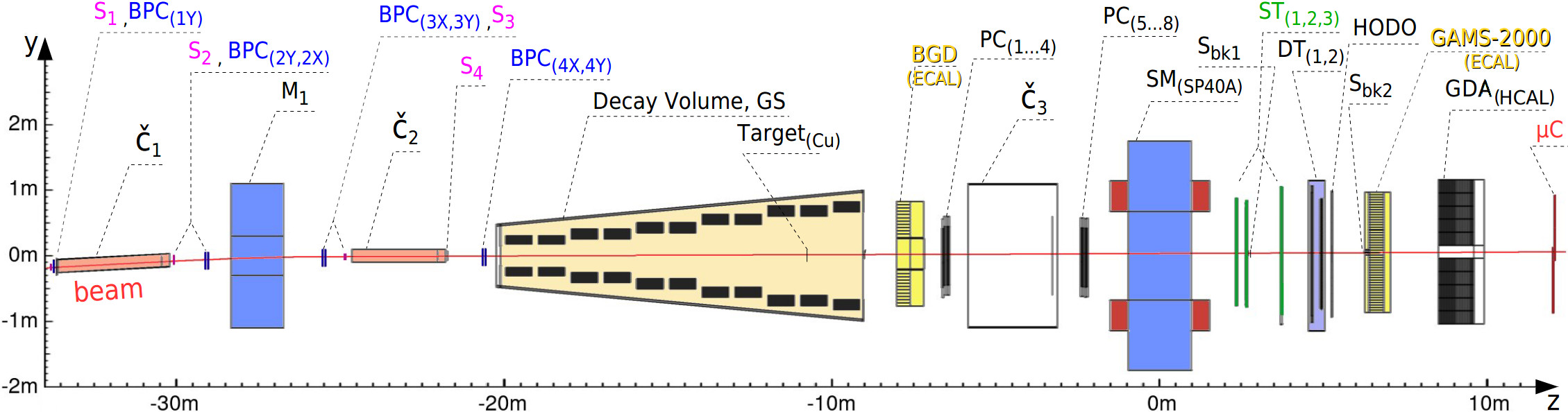}
\caption{\label{FigOkaSetup} \descrfnt Schematic elevation view of the OKA setup.}
\end{figure}

The OKA setup  uses two magnetic spectrometers placed along the beam line with an 11~m long {\small Decay Volume (DV)} 
in-between, see Fig.~\ref{FigOkaSetup}. The {\small DV} is filled with helium and contains 11 rings of guard system ({\small GS}) made of Lead-Scintillator sandwiches
at its inner surface. 
The guard system is complemented by an electromagnetic calorimeter {\small BGD} with a wide central opening.

Momenta measurement of the beam and of charged decay products is done with a help of {\small M$_{1}$} and wide aperture 200$\times$140 cm$^2$ {\small M$_{2}$} magnets.
The magnets are surrounded by a set of beam proportional chambers {\small BPC} and a set of large apperture proportional chambers {\small PC}, 
straw tubes {\small ST} and drift tubes {\small DT}. 
A matrix hodoscope {\small HODO} is used to improve the time resolution and to link $x$--$y$ projections of a track.

At the end of the setup there are the main electromagnetic calorimeter {\small GAMS$_{(ECAL)}$}, made of $\sim2300$ 18$X_{0}$ lead glass blocks, 
the hadron calorimeter, {\small GDA$_{(HCAL)}$}, made of 100 $5\lambda$ iron-scintillator sandwiches 
and a muon counter {\small {$\mu$}C}, consisting of four partially overlapping 1$\times1${\hspace{0.5pt}}m$^2$ scintillator plates at the end of the setup.
The data acquisition system of the OKA setup \cite{OKA_DAQ} operates at $\sim25$ kHz event rate with the mean event size of $\sim4$ kByte.
More details on the OKA setup are available in our previous publications \cite{OKA_Ke3_JETP, OKA_KmuHnu_EPJC}.

To calculate the detection efficiency and the detector response, the Monte Carlo (MC) simulation 
is performed using a software package based on the Geant-3.21 library \cite{Geant321},
which includes a realistic description of the setup. 
The trigger used in the experiment is also included in the simulation. 
The MC events are passed through the full OKA reconstruction procedure, same as for the experimental data.
To estimate the background, samples of MC events are used for five main decay channels of charged kaon with $\pi^{0}$ in the final state:
($K^{+}$$\to$~~$\pi^{+}\pi^{0}$, $\pi^{+}\pi^{0}\gamma$, $\pi^{+}\pi^{0}\pi^{0}$, $\pi^{0}\mu^{+}\nu$, and $\pi^{0}e^{+}\nu$). 
They are mixed according to their branching fractions.
Monte Carlo events use weights proportional to the square of the corresponding matrix elements according to the PDG \cite{PDG}.
The generated MC statistics is $\sim$ 8 times larger than the data sample recorded in the experiment.

A dedicated MC is produced for the process of interest $K^{+} \to \pi^{+} \pi^{0} \overline{\gamma}$.
It uses realistic matrix element from  \cite{jySu_jTandean_EPJC80_2020_9_824}, see section \ref{sectIntro}.

\section{The data set and the event selection}\label{sectDataEventSel}
A following trigger: ${\tt Tr_{GAMS}=}$ ${\tt S_{1}{\times}S_{2}{\times}S_{3}{\times}S_{4}{\times}\overline{\check{C}}_{1}{\times}\check{C}_{2}{\times}\overline{S}_{bk} }\times (E_{GAMS}>2.5$~GeV$)$
is used to collect data for the present analysis.
The coincidence of four scintillation counters  ({\small {S}$_{1}$, {S}$_{2}$, {S}$_{3}$, and {S}$_{4}$}) selects a beam particle.
The final selection of charged kaons is done with a combination of two threshold Cherenkov counters ({\small \v{C}$_{1}$} sees pions, {\small \v{C}$_{2}$} sees pions and kaons).
An anti-coincidence with two scintillation counters ({\small S$_{bk1}$ or S$_{bk2}$}) located on the beam axis behind the {\small M$_{2}$} magnet is used to avoid recording the events with undecayed beam particles.
The cut on the energy deposition in the {\small GAMS} e.m.~calorimeter suppresses  the dominating $K^{+} \to \mu^{+} \nu$ decay and enhances decays with photons and electrons.

Two sequential sets of data with a beam momentum of 17.7 GeV/c recorded by the OKA collaboration in 2012 and 2013 ({\tt run14} and {\tt run15}) are used in the analysis to search for the massless dark photon.
The total number of $\sim$ $2.65\times 10^{9}$ events with ${\tt Tr_{GAMS}}$ are logged. 
One of the tasks during data taking in {\tt run14} was a study of the $K^{+}$Cu$\to K^{+} \pi^{0}$Cu reaction \cite{BurtovoyCoherrentKPI0},
for that a thin copper target was installed inside the {\small DV} near its exit window for the half time of the run. 
A dedicated Monte-Carlo simulation (MC) for that period was performed. 

As the first step of the reconstruction, events with kaon decay with a single charged track in the final state are selected, 
with the decay vertex inside the {\small DV} and with reconstructed momenta in both magnetic spectrometers.
About $8\times 10^{8}$ events are selected.

\subsection{Event selection}
\vspace{-3.0pt}
The event selection for the decay $K^{+}\to\pi^{+}\pi^{0}{\overline{\gamma}}$ starts from the reconstruction of $\pi^{+}\pi^{0}$ final state.
A single beam track and a single secondary track with a decay angle $>4$~mrad and with a vertex matching (distance of closest approach) below $1.25$~cm is required.
A moderate chi-square cut for the charged track quality is applied.
To clean the tracking sample, it is required that no extra track segments behind the {\small M$_{2}$} magnet are found.
The vertex position is required to be inside the {\small DV} within two sigma margins from its entrance and exit windows,
and outside the target position for the run periods in which the target was used.
The beam particle momentum is required within a range of $17.0<p_{beam}<18.6$~GeV/c. 
The number of showers in the {\small GAMS} calorimeter not associated with the track must be equal to 2. 
For these events, the $\pi^{0}$ identification is done by invariant mass selection: $|m_{\gamma\gamma}-m_{\pi^{0}}|<15$~MeV/c$^{2}$
for both gammas with $E_{\gamma}>0.3$~GeV.
After those selections the statistics is dominated by $K^{+}\to\pi^{+}\pi^{0}$ decay. 
We obtain $\sim 3\times 10^{7}$ $K^{+}\to\pi^{+}\pi^{0}$ events ($16.9\times10^{6}$ and $11.4\times10^{6}$ in {\tt run14} and {\tt run15} correspondingly).
At this stage, the normalization of MC statistics to the data is performed.

\section{Signal selection, background analysis}
First, a cut $E_{mis}$$=$$(E_{K^{+}} - E_{\pi^{+}} - E_{\pi^{0}})$$>$$2.7$~GeV on the missing energy is applied in order to disentangle the $K^{+}\to\pi^{+}\pi^{0}\overline{\gamma}$ signal 
from its main background $K^{+}\to\pi^{+}\pi^{0}$.

\begin{figure}[!ht]
\centering
\includegraphics[width=0.99\textwidth]{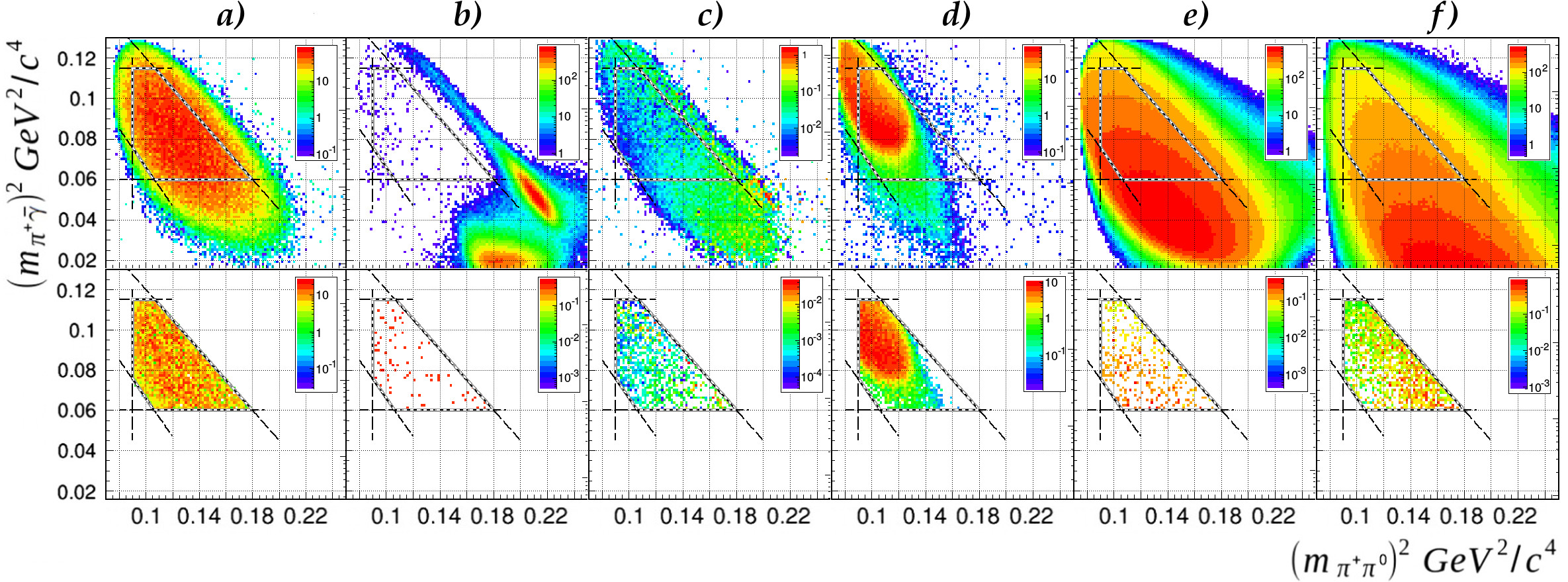}
\caption{\label{MC_DalitzPlots_DP_and_5_backgrounds} 
\descrfnt
 The $(m_{\pi^{+}\overline{\gamma}})^2$~vs.~$(m_{\pi^{+}\pi^{0}})^2$ Dalitz plots for reconstructed MC events 
 from the signal process {\it (a)} and from its main background processes: 
 {\it (b)} $K^{+}\to\pi^{+}\pi^{0}$,
 {\it (c)} $K^{+}\to\pi^{+}\pi^{0}\gamma$,
 {\it (d)} $K^{+}\to\pi^{+}\pi^{0}\pi^{0}$, 
 {\it (e)} $K^{+}\to\mu^{+}\nu\pi^{0}$ and 
 {\it (f)} $K^{+}\to e^{+}\nu\pi^{0}$.
 The upper row shows these distributions after the cut on the missing energy $E_{mis}>2.7$~GeV.
 The inner area bounded by the dashed lines is selected for the signal.
 The bottom row shows the distributions after the final selections. Note the logarithmic scale of the plot density.
}
\end{figure}

Figure \ref{MC_DalitzPlots_DP_and_5_backgrounds}
demonstrates MC Dalitz plots for the signal and for the main background channels.
To accept the signal of $K^{+}\to\pi^{+}\pi^{0}{\overline{\gamma}}$ 
and sufficiently suppress the background we select an area inside the dashed region  
as seen in Fig.~\ref{MC_DalitzPlots_DP_and_5_backgrounds}. 
Despite the $K^{+}\to\pi^{+}\pi^{0}\pi^{0}$ looks fully accepted, it does not affect signal region in the $(m_{mis})^2$ plot 
since it is well separated from it as it is seen in Fig.~\ref{mm2distrib_r14r15_DarkPhotonSearch}. 
This is not the case for $K^{+}\to\mu^{+}\nu\pi^{0}$ and $K^{+}\to e^{+}\nu\pi^{0}$ which are considerably
populating the region of the signal with $(m_{mis})^2 \approx 0$.
The area selected for the signal strongly minimizes possible inputs from $K^{+}\to\pi^{+}\pi^{0}$ and $K^{+}\to\pi^{+}\pi^{0}\gamma$
due to drastically different matrix elements. 

To further suppress $K^{+}\to\mu^{+}(e^{+})\nu\pi^{0}$ 
we use the $\pi^{+}$/$\mu^{+}$/$e^{+}$ separation by means of the muon veto and the calorimetry. 
The muon veto requires absence of a signal from the muon counter crossed by the extrapolated track. 
The requirements for the hadron shower in {\small GAMS} include the track matching with a {\small GAMS} cluster within $\Delta R<8$~cm;
the $E/p<0.65$ cut to avoid showers from $e^{+}$,
while preserving the "early" hadron ($\pi^{+}$) showers
for which an "anti" cut on the cluster fit quality with an electromagnetic shower $\chi^{2}_{em}>0.4$ is required, 
which is also suited to reject muon clusters.
The resolution of hadronic calorimeter {\small HCAL} in the energy range of interest is insufficient for strong $\pi^{+}$/$\mu^{+}$ separation
that is why it was decided to omit HCAL from PID.

At the end of the procedure it is ensured that the ($\vec{p}_{\overline{\gamma}}$) missing momentum direction is crossing the {\small GAMS} acceptance.
Finally, the veto on the total energy deposition in the GS rings and {\small BGD} is applied, i.e.~$\Sigma E_{GS}<100$~MeV 
and $\Sigma E_{BGD}<200$~MeV.

\section{Signal search and upper limit settings}\label{sectSignalSearch}
Since the massless dark photon would manifest itself as a peak at $(m_{mis})^2=0$,  
a search for the signal peak around $m^{2}_{mis}\cong 0$~GeV${}^{2}$/${c^{4}}$ is performed, for $m^2_{\overline{\gamma}} = m^2_{mis} = ({\textrm p}_{K^{+}} - {\textrm p}_{\pi^{+}} - {\textrm p}_{\pi^{0}})^{2}$, 
where ${\textrm p}_{K^{+}}$, ${\textrm p}_{\pi^{+}}$ and ${\textrm p}_{\pi^{0}}$ are four-momenta of the corresponding particles. 
The shape of the MC signal is well described by the Gaussian, the resolution in $(m_{mis})^2$ is $\sim($50~MeV/$c^{2}$$)^{2}$, see table~\ref{refTable1}.

\begin{figure}[!ht]
\centering
\includegraphics[width=1.00\textwidth]{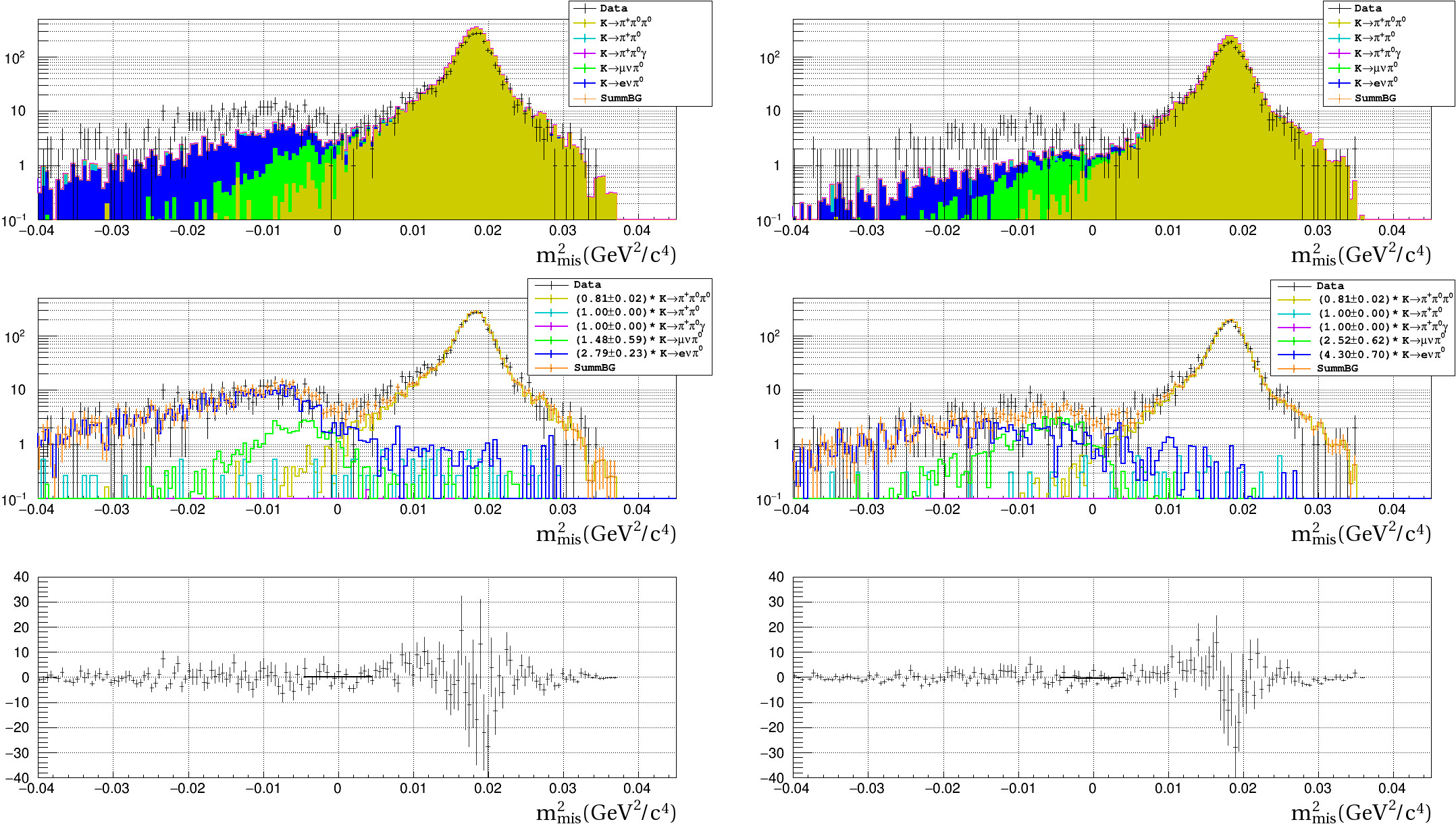}
\caption{\label{mm2distrib_r14r15_DarkPhotonSearch} 
   \descrfnt
   {\tt{Upper plots}}: the resulting $m^{2}_{mis}$-distribution for the {\tt run14} (left) and {\tt run15} (right), 
   the stack plot is used, note the {\it log-y} scale. 
   Black points with errors correspond to the experimental data.
   The background channels and their sum obtained from the MC are marked with colors.
   In the {\tt{middle plots}} they are shown after the tuning of the relative magnitudes,
   the corresponding scaling factors are depicted with their fit errors.
   The sum of the background channels is indicated with orange.
   {\tt{Bottom plots}} demonstrate the difference between the experiment and the sum of the tuned background processes.
   The result of the signal fit at $m^{2}_{mis}=0$ is shown with black line.
}
\end{figure}

The resulting $m^{2}_{mis}$-distributions for the data and for the main background processes are shown in Fig.~\ref{mm2distrib_r14r15_DarkPhotonSearch}. 
Due to the strong suppression cuts (up to $\sim 10^{5}$), the efficiency estimates for the background processes are known with noticeable errors, so a (maximum likelihood) fit
in which the efficiencies of the main background processes are allowed to vary is used to tune the background model to the experimental data.
The background fit is performed inside a wide region of $(-0.025<m^{2}_{mis}<0.025)$~GeV${}^{2}/c^{4}$.
Since $K^{+}\to\pi^{+}\pi^{0} (\gamma)$ peak at the same position as the signal, $m^{2}_{mis}=0$, they are excluded from the fit to prevent an absorption of signal events. 
As a result of the fit we obtain a reasonable description of the background in the wide $m^{2}_{mis}$-mass range, 
the discrepancy mainly concerns the $\pi^{0}$-mass region of the $K^{+} \to \pi^{+} \pi^{0} \pi^{0}$ background.

After that a maximum likelihood (ML) fit of the data using signal Gaussian shape as determined from the MC is done on top of the tuned background. 
The fit is repeated at three points near $m^{2}_{mis}=0$ to include possible systematics in the missing mass scale. 
A 5\% uncertainty for the signal width is allowed in the fit to include a possible inaccuracy in describing the signal width by the MC.
In the absence of a statistically significant signal, the upper limit is calculated for the number of signal events.
We follow the method used in \cite{oTchikilevISTRA2004}, in which the fit procedure for the signal is not bound to positive values only (to avoid a possible bias).
The one-sided upper limit for the number of signal events corresponding to the confidence level (CL) of 90\% is constructed as $N_{UL@90{\%}CL} = max(N_{\overline{\gamma}},0)+1.28\cdot\sigma_{N_{\overline{\gamma}}}$,
where $N_{\overline{\gamma}}$ and $\sigma_{N_{\overline{\gamma}}}$ are evaluated from the ML fit, being the number of signal events and its error.

In contrast to the Feldman-Cousins \cite{FeldmanCousinsMethod} method, where only the central region of the signal is considered, 
this approach is chosen because it allows one to take advantage of knowledge of the signal shape in a wider range.
The obtained parameters, $N_{\overline{\gamma}}$ and $\sigma_{N_{\overline{\gamma}}}$, are used to perform a statistical combination of the two runs.

\begin{figure}[!ht]
\centering
\includegraphics[width=0.99\textwidth]{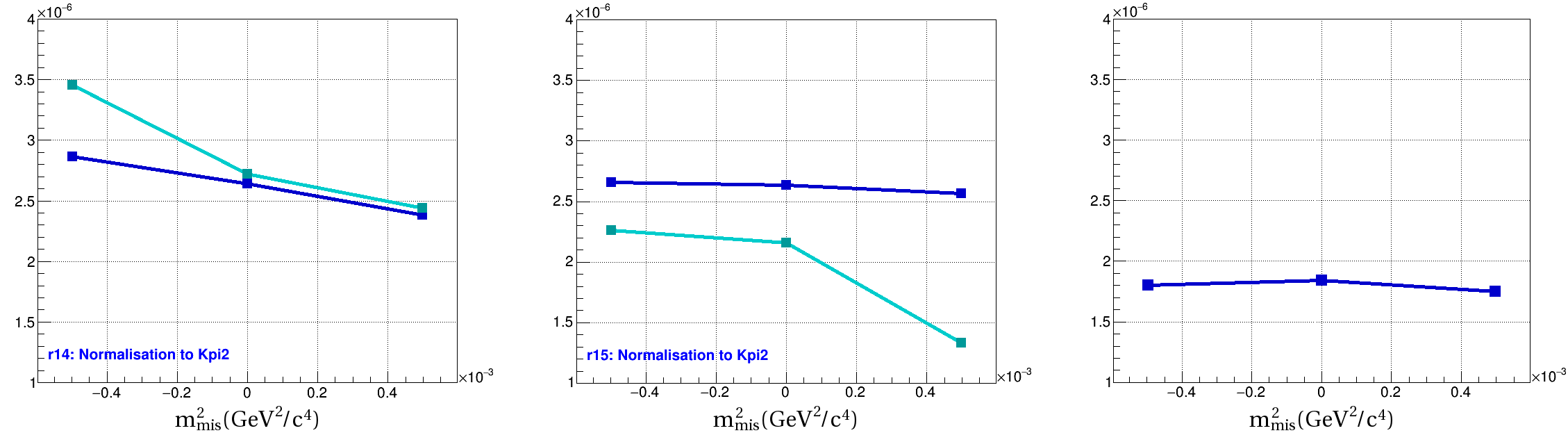}
\caption{\label{resultsDP_UL_CL90_r14r_15_MergedML} 
   \descrfnt
   Results for the $K^{+}\to\pi^{+}\pi^{0}\overline{\gamma}$ branching fraction upper limit at the 90\% CL for run 14, ({\tt left plot}) run 15 ({\tt middle plot}) for three points near $m_{mis}=0$ 
   obtained with two methods: 
   maximum likelihood (ML) fit shown with blue boxes 
   and Feldman-Cousins method (applied within the window of 1.2 sigma, with the number of events corrected for the signal full width) shown with green-blue boxes. 
   Statistical combination of ML-fit results for two runs is shown in the {\tt right plot}.
}
\end{figure}

The upper limits for the branching ratio are calculated using relative normalization to the $K^{+}\to\pi^{+}\pi^{0}$ (aka $K\pi2$) decay \cite{PDG}.
The branching for the signal decay is calculated as follows:\\
$Br(K^{+}$$\to$$\pi^{+}\pi^{0}{\overline{\gamma}})=Br(K\pi2)\times (N_{\overline{\gamma}}/\varepsilon_{\overline{\gamma}})\times(\varepsilon_{K\pi2}/N_{K\pi2})$, 
where $N_{K\pi2}$ is the number of events\footnote{When $K\pi2$ is selected for the normalization, 
      the cut in the Dalitz plot (see Fig.~\ref{MC_DalitzPlots_DP_and_5_backgrounds}) is not applied, 
      while the cut on $E_{mis}$ is inverted.} 
      for $K^{+}\to\pi^{+}\pi^{0}$,
      $\varepsilon_{K\pi2}$ 
      and $\varepsilon_{\overline{\gamma}}$ are the $K\pi2$ and the signal efficiency estimated with the MC, see table~\ref{refTable1}; 
      the $N_{\overline{\gamma}}$ is taken from the fit. 
The results for each run and for the combined result are shown in Fig.~\ref{resultsDP_UL_CL90_r14r_15_MergedML}.
The combined UL on $Br(K^{+}$$\to$$\pi^{+}\pi^{0}{\overline{\gamma}})<1.8\times10^{-6}$ is reached.
The main sources of systematics related to the trigger, track quality, and particle identification cancel out in this approach.

\begin{table*}[htbp]
\begin{minipage}{0.50\linewidth}
 \begin{center}
 \begin{tabular}{|c|c|c|}\hline
                                                 & {\tt run14}                 & {\tt run15}          \\ \hline
 $\varepsilon_{K\pi2}$                           & $3.1\%$                     & $3.3\%$              \\ \hline
 $N_{K\pi2}$                                     & $4.1\times10^6$             & $2.7\times10^6$      \\ \hline
 $\varepsilon_{\overline{\gamma}}$               & $0.67\%$                    & $0.72\%$             \\ \hline
 $\sigma_{\overline{\gamma}}$, {\tiny GeV${}^{2}/c^{4}$} & $2.4\times10^{-3}$  & $2.3\times10^{-3}$   \\ \hline
 \end{tabular}
 \end{center}
\end{minipage}
\begin{minipage}{0.450\linewidth}
\caption{
  \descrfnt
  Parameters used for DP branching ratio calculation. Quoted efficiencies include acceptance and are normalized to kaon decays inside the {\small DV}.
  \label{refTable1}
  \vspace{0.5cm}
}
\end{minipage}
\end{table*}

\subsection{Estimates of systematic error}
The systematic error in $\varepsilon_{\overline{\gamma}}$ is about 5\%, 
while the systematic error of the $(\varepsilon_{K\pi2}/N_{K\pi2})$ ratio is estimated as $3.5\%$, the corresponding statistical errors are negligible.

A systematic error of {11\%} is derived from the difference between the $K^{+}\to\pi^{+}\pi^{0}$ and $K^{+}\to\pi^{+}\pi^{0}\pi^{0}$ (with lost $\pi^{0}$, which is detected by the missing mass spectrum) normalization alternatives.
For the $K^{+} \to \pi^{+}\pi^{0}\pi^{0}$ case, the veto ({\small GS}) cut is removed to avoid suppression of the second (escaping) $\pi^{0}$, while (in contrast to the $K^{+}\to\pi^{+}\pi^{0}$) the $E_{mis}$ cut is kept unchanged. 

\begin{figure}[!ht]
\centering
\includegraphics[width=0.50\textwidth]{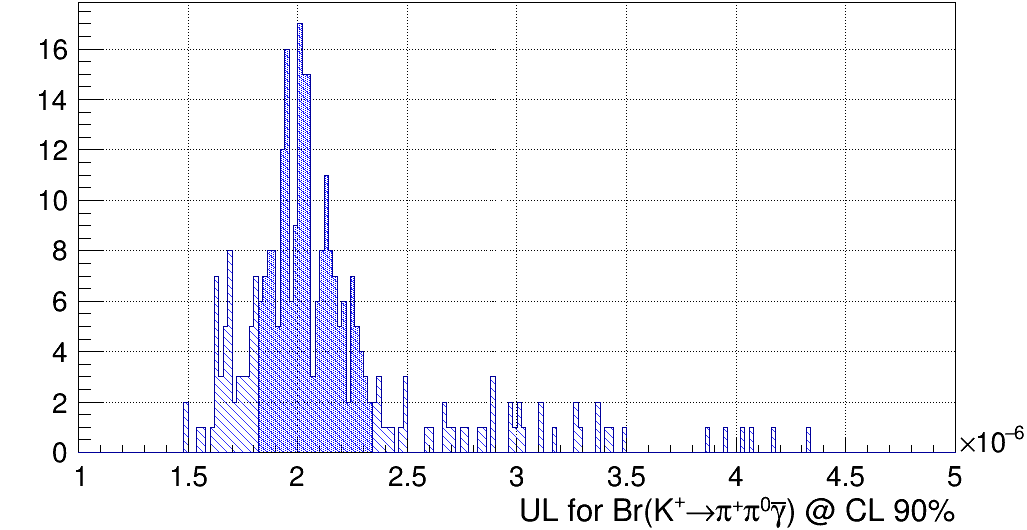}
\caption{\label{UL90percentCL_DarkPhoton_100reanalysis} 
   \descrfnt
   Distribution for the upper limit (90\% CL) on $Br$($K^{+}\to\pi^{+}\pi^{0}\overline{\gamma}$) for 100 standard analysis procedures 
   (described in section \ref{sectSignalSearch}) with randomly chosen cutoffs in each of the analyses. 
   The results obtained at three points near $m^{2}_{mis}=0$ are shown together.
   The regions, which contain 16\% of events from each side are highlighted in light shading.
}
\end{figure}

To account for a possible systematic error due to the selection criteria, we repeated a hundred standard analyses described earlier in this paper,
where a set of the main cuts was randomly chosen,                    within a window of $\pm \sigma$ 
(where $\sigma$ is either the resolution of a cut variable, estimated from the MC, or $\sim$ 15\% for threshold cuts) around the original values.
The distribution of the resulting UL is shown in Fig.~\ref{UL90percentCL_DarkPhoton_100reanalysis}.
The  most probable value upper limit is considered as the final result, 
while the systematic error of the upper limit is estimated by rejecting the tails 
of 16\% from each side of the distribution (keeping inside 68\% of the distribution), which corresponds to UL on $Br(K^{+}$$\to$$\pi^{+}\pi^{0}{\overline{\gamma}}) < 2.0{^{+0.36}_{-0.19}}\times 10^{-6}$. 
This leads to UL at the 90\% CL on $Br(K^{+}$$\to$$\pi^{+}\pi^{0}{\overline{\gamma}}) < 2.0{^{+0.44}_{-0.32}}(syst.)\times 10^{-6}$, if other systematic errors are added in quadrature.

\section{Conclusions}
The obtained upper limit $Br(K^{+}$$\to$$\pi^{+}\pi^{0}{\overline{\gamma}})<2.0{^{+0.4}_{-0.3}}(syst.)\times10^{-6}$ is weaker than that 
from the theoretical model \cite{mFabbrichesi_etal_PRL119_031801}, $Br(K^{+}$$\to$$\pi^{+}\pi^{0}{\overline{\gamma}})<1.7\times10^{-7}$,
although it is better or comparable with that of \cite{jySu_jTandean_EPJC80_2020_9_824}.
Using the relationship from \cite{jySu_jTandean_EPJC80_2020_9_824} it leads to the limit on the parameters 
of the FCNC $s \to d \overline{\gamma}$ matrix element: $|\mathbb{C}|^{2} + |\mathbb{C}_{5}|^{2} < 1.8\times10^{-16}$~${\textrm{GeV}^{-2}}$.

The limit on the $(|\mathbb{C}|^2 + |\mathbb{C}_{5}|^2)$ can be compared with that obtained from 
the hyperon decays $\Lambda \to n \overline{\gamma}$, $\Sigma^{+} \to p \overline{\gamma}$ and others 
with the same FCNC transition mechanism  ($s \to d  \overline{\gamma}$) behind. 
In line with this, several experiments are to be mentioned.

The BESIII Collaboration performed a search for the dark photon in the hyperon decay $\Sigma^{+} \to p + invisible$ 
and obtained an upper limit of $3.8\times10^{-5}$ (at the 90\% CL) on the branching fraction 
for the massless dark photon \cite{BESIII_collaboration_Sigma_To_Proton_Invisible}.
This can be converted into a limit $|\mathbb{C}|^{2} + |\mathbb{C}_{5}|^{2}<2.08\times10^{-16}$~${\textrm{GeV}^{-2}}$, according to \cite{jySu_jTandean_EPJC80_2020_9_824}.

One should also mention an impressive limit $[(\mathtt{Re}\mathbb{C})^{2} + (\mathtt{Im}\mathbb{C}_{5})^{2}]<6\times10^{-20}$~${\textrm{GeV}^{-2}}$ which can be obtained (see eq.~17 in \cite{jySu_jTandean_EPJC80_2020_9_824})
from the KOTO \cite{tWu_KOTO_KLtoGammaDP_PoS2024} result $Br(K^{0}_{L}\to \gamma \overline{\gamma})<3.47\times10^{-7}$. 

\subsection*{Funding}\vspace{-7pt}
The work was supported by the RSCF grant {\it{N\textsuperscript{\underline{\scriptsize o}}}}22-12-00051-$\Pi$.

\section*{Acknowledgments}\vspace{-7pt}
We express our gratitude to our colleagues in the accelerator department for the good performance of the U-70 during data taking; 
to colleagues from the beam department for the stable operation of the 21K beamline, including RF-deflectors, and to colleagues 
from the engineering physics department for the operation of the cryogenic system of the RF deflectors.


\end{document}